\documentclass[runningheads]{llncs}
\usepackage[T1]{fontenc}
\usepackage{graphicx}
\usepackage{float}
\usepackage{gensymb}
\usepackage{amsmath}
\usepackage{hyperref}
\usepackage{color}

\usepackage[scale=0.35, angle=0, color={[gray]{0.3}}, vpos=0.06\paperheight, text={This version of the contribution has been accepted for publication, after peer review, \\ but is not the Version of Record and does not reflect post-acceptance improvements, \\ or any corrections. The Version of Record is available online at: \\ https://doi.org/10.1007/978-3-031-12053-4\_65.}]{draftwatermark}

\begin{document}

\title{FCN-Transformer Feature Fusion for Polyp Segmentation}

\author{Edward Sanderson \and\orcidID{0000-0002-3794-5513} \and \\
Bogdan J. Matuszewski\orcidID{0000-0001-7195-2509}}
\authorrunning{E. Sanderson, B. J. Matuszewski}

\institute{Computer Vision and Machine Learning (CVML) Group, University of Central Lancashire, Preston, UK \\
\email{\{esanderson4, bmatuszewski1\}@uclan.ac.uk}\\
\url{https://www.uclan.ac.uk/research/activity/cvml}}

\maketitle
\begin{abstract}

Colonoscopy is widely recognised as the gold standard procedure for the early detection of colorectal cancer (CRC). Segmentation is valuable for two significant clinical applications, namely lesion detection and classification, providing means to improve accuracy and robustness. The manual segmentation of polyps in colonoscopy images is time-consuming. As a result, the use of deep learning (DL) for automation of polyp segmentation has become important. However, DL-based solutions can be vulnerable to overfitting and the resulting inability to generalise to images captured by different colonoscopes. Recent transformer-based architectures for semantic segmentation both achieve higher performance and generalise better than alternatives, however typically predict a segmentation map of $\frac{h}{4}\times\frac{w}{4}$ spatial dimensions for a $h\times w$ input image. To this end, we propose a new architecture for full-size segmentation which leverages the strengths of a transformer in extracting the most important features for segmentation in a primary branch, while compensating for its limitations in full-size prediction with a secondary fully convolutional branch. The resulting features from both branches are then fused for final prediction of a $h\times w$ segmentation map. We demonstrate our method's state-of-the-art performance with respect to the mDice, mIoU, mPrecision, and mRecall metrics, on both the Kvasir-SEG and CVC-ClinicDB dataset benchmarks. Additionally, we train the model on each of these datasets and evaluate on the other to demonstrate its superior generalisation performance.

Code available: \url{https://github.com/CVML-UCLan/FCBFormer}

\end{abstract}

\section{Introduction}

Colorectal cancer (CRC) is a leading cause of cancer mortality worldwide; e.g., in the United States, it is the third largest cause of cancer deaths, with 52,500 CRC deaths predicted in 2022 \cite{bog1}. In Europe, it is the second largest cause of cancer deaths, with 156,000 deaths in 27 EU countries reported in 2020 \cite{bog2}.

Colon cancer survival rate depends strongly on an early detection. It is commonly accepted that most colorectal cancers evolve from adenomatous polyps \cite{bog3}. Colonoscopy is the gold standard for colon screening as it can facilitate detection and treatment during the same procedure, e.g., by using the resect-and-discard and diagnose-and-disregard approaches. However, colonoscopy has some limitations; e.g., It has been reported that between 17\%-28\% of colon polyps are missed during colonoscopy screening procedures \cite{bog4,bog5}. Importantly, it has been assessed that improvement of polyp detection rates by 1\% reduces the risk of CRC by approximately 3\% \cite{bog6}. It is therefore vital to improve polyp detectability. Equally, correct classification of detected polyps is limited by variability of polyp appearance and subjectivity of the assessment. Lesion detection and classification are two tasks for which intelligent systems can play key roles in improving the effectiveness of the CRC screening and robust segmentation tools are important in facilitating these tasks.

To improve on the segmentation of polyps in colonoscopy images, a range of deep learning (DL) -based solutions \cite{resunet++,pranet,colonsegnet,ddanet,uacanet,hardnet,transfuse,caranet,msrfnet,ssformer} have been proposed. Such solutions are designed to automatically predict segmentation maps for colonoscopy images, in order to provide assistance to clinicians performing colonoscopy procedures. These solutions have traditionally used fully convolutional networks (FCNs) \cite{unet,unet++,resunet++,guo1,guo2,doubleunet,guo3,colonsegnet,hardnet,msrfnet}. However, transformer-based architectures \cite{dpt,segformer,pvt,pvtv2,ssformer} have recently become popular for semantic segmentation and shown superior performance over FCN-based alternatives. This is likely a result of the ability of transformers to efficiently extract features on the basis of a global receptive field from the first layers of the model through global attention. This is especially true in generalisability tests, where a model is trained on one dataset and evaluated on another dataset in order to test its robustness to images from a somewhat different distribution to that considered during training. Some studies have also combined FCNs and transformers/attention mechanisms \cite{pranet,ddanet,uacanet,transunet,transfuse,caranet} in order to combine their strengths in a single architecture for medical image segmentation, however these hybrid architectures do not outperform the highest performing FCN-based and transformer-based models in this task, notably MSRF-Net \cite{msrfnet} (FCN) and SSFormer \cite{ssformer} (transformer). One significant limitation of most the highlighted transformer-based architectures is however that the predicted segmentation maps of these models are typically of a lower resolution than the input images, i.e. are not full-size. This is due to these models operating on tokens which correspond to patches of the input image rather than pixels.

\begin{figure}[H]
\makebox[\textwidth][c]{
\includegraphics[height=12.5cm]{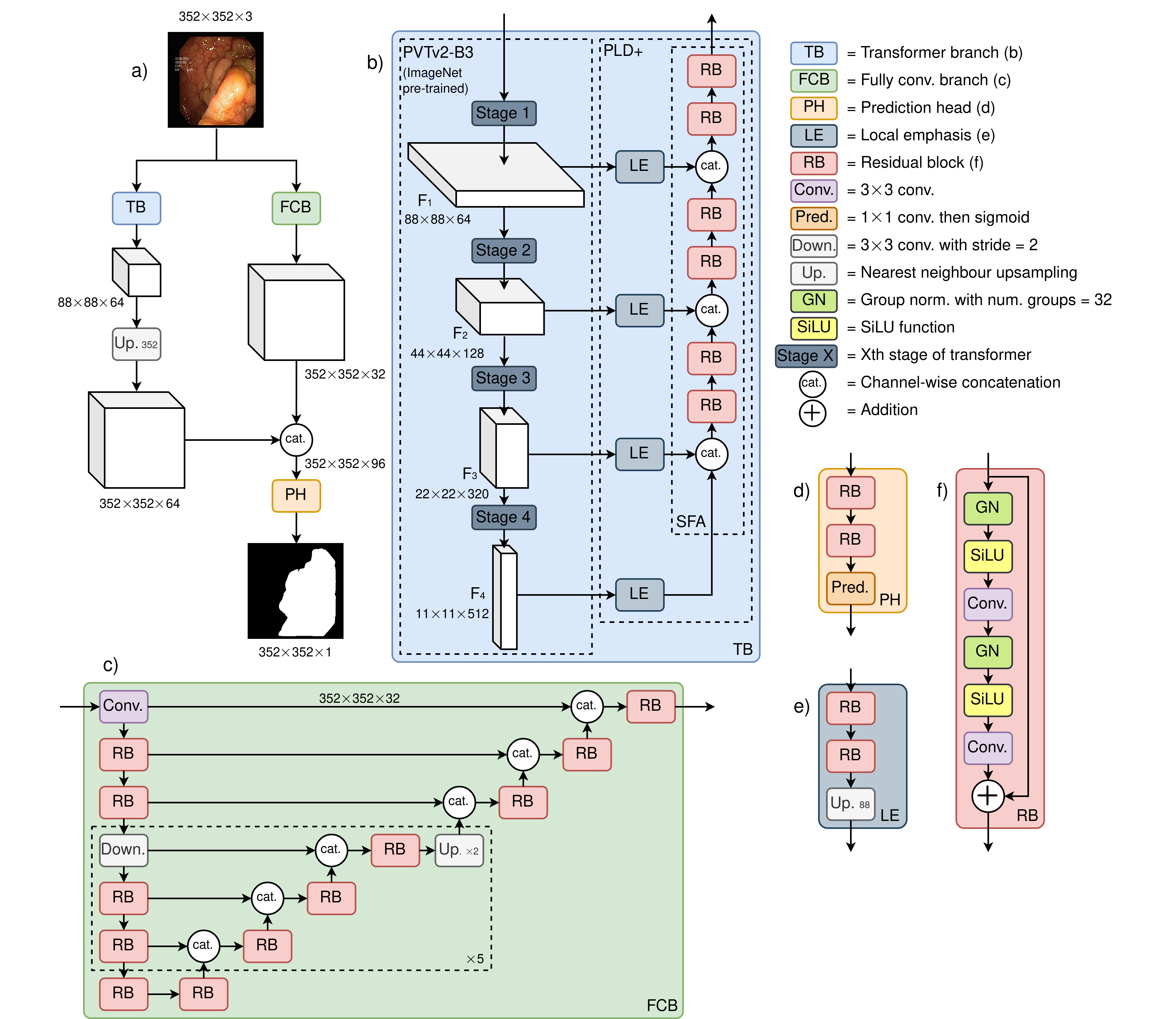}}
\caption{The architectures of a) FCBFormer, b) the transformer branch (TB), c) the fully convolutional branch (FCB), d) the prediction head (PH), e) the improved local emphasis (LE) module, f) the residual block (RB).} \label{fig:FCBformer}
\end{figure}

In this paper, we propose a new architecture for polyp segmentation in colonoscopy images which combines FCNs and transformers to achieve state-of-the-art results. The architecture, named the Fully Convolutional Branch-TransFormer (FCBFormer) (Fig. \ref{fig:FCBformer}a), uses two parallel branches which both start from a $h\times w$ input image: a fully convolutional branch (FCB) which returns full-size ($h\times w$) feature maps; and a transformer branch (TB) which returns \makebox[\linewidth][s]{reduced-size ($\frac{h}{4}\times\frac{w}{4}$) feature maps. The output tensors of TB are then ups-} \par

\noindent ampled to full-size, concatenated with the output tensors of FCB along the channel dimension, before a prediction head (PH) processes the concatenated tensors into a full-size segmentation map for the input image. Through the use  of the ImageNet \cite{imagenet} pre-trained pyramid vision transformer v2 (PVTv2) \cite{pvtv2} as an image encoder, we encourage the model to extract the most important features for segmentation in TB. We then randomly initialise FCB to encourage extraction of the features required for processing outputs of TB into full-size segmentation maps. TB largely follows the structure of the recent SSFormer \cite{ssformer} which predicts segmentation maps of $\frac{h}{4}\times\frac{w}{4}$ spatial dimensions, and which achieved the current state-of-the-art performance on polyp segmentation at reduced-size. However, we update the SSFormer architecture with a new progressive locality decoder (PLD) which features improved local emphasis (LE) and stepwise feature aggregation (SFA). FCB then takes the form of an advanced FCN architecture, composed of a modern variant of residual blocks (RBs) that include group normalisation \cite{groupnorm} layers, SiLU \cite{gelu} activation functions, and convolutional layers, with a residual connection \cite{resnet,highwaynets}; in addition to dense U-Net style skip connections \cite{unet}. PH is then composed of RBs and a final pixel-wise prediction layer which uses convolution with 1$\times$1 kernels. On this basis, we achieve state-of-the-art performance with respect to the mDice, mIoU, mPrecision, and mRecall metrics on the Kvasir-SEG \cite{kvasir} and CVC-ClinicDB \cite{cvc} datasets, and on generalisability tests where we train the model on one Kvasir-SEG and evaluate it on CVC-ClinicDB, and vice-versa.

The main novel contributions of this work are therefore:
\begin{enumerate}
\item The introduction of a simple yet effective approach for FCNs and transformers in a single architecture for dense prediction which, in contrast to previous work on this, demonstrates advantages over these individual model types through state-of-the-art performance in polyp segmentation.
\item The improvement of the progressive locality decoder (PLD) introduced with SSFormer \cite{ssformer} for decoding features extracted by a transformer encoder through residual blocks (RBs) composed of group normalisation \cite{groupnorm}, SiLU activation functions \cite{groupnorm}, convolutional layers, and residual connections \cite{resnet}.
\end{enumerate}

The rest of this paper is structured as follows: we first define the design of FCBFormer and its components in Section \ref{sec:FCBFormer}; we then outline our experiments in terms of the implementation of methods, the means of evaluation, and our results, in Section \ref{sec:experiments}; and in Section \ref{sec:conclusion} we give our conclusion.

\section{FCBFormer}\label{sec:FCBFormer}
\subsection{Transformer branch (TB)\label{TB}}
\begin{figure}[htp]
\makebox[\textwidth][c]{
\includegraphics[height=8cm]{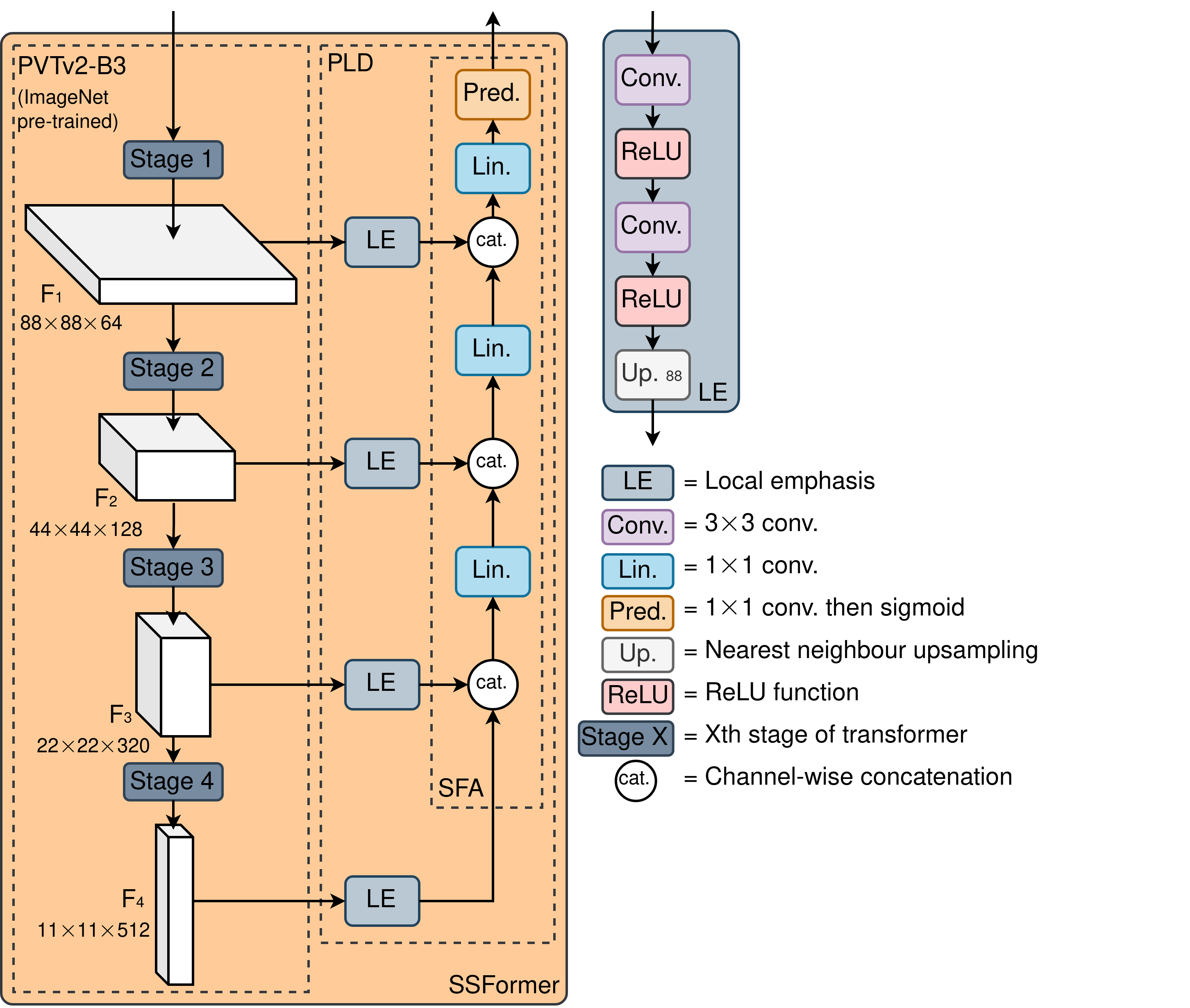}}
\caption{The architecture of our implementation of SSFormer.} \label{fig:SSformer}
\end{figure}

The transformer branch (TB) (Fig. \ref{fig:FCBformer}b) is highly influenced by the current state-of-the-art architecture for reduced-size polyp segmentation, the SSFormer \cite{ssformer}. Our implementation of SSFormer, as used in our experiments, is illustrated in Fig. \ref{fig:SSformer}. This architecture uses an ImageNet \cite{imagenet} pre-trained pyramid vision transformer v2 (PVTv2) \cite{pvtv2} as an image encoder, which returns a feature pyramid with 4 levels that is then taken as the input for the progressive locality decoder (PLD). In PLD, each level of the pyramid is processed individually by a local emphasis (LE) module, in order to address the weaknesses of transformer-based models in representing local features in the feature representation, before fusing the locally emphasised levels of the feature pyramid through stepwise feature aggregation (SFA). Finally, the fused multi-scale features are used to predict the segmentation map for the input image.

PLD takes the tensors returned by the encoder, with a number of channels defined by PVTv2, and changes the number of channels in the first convolutional layer in each LE block to 64. Each subsequent layer, except channel-wise concatenation and the prediction layer, then returns the same number of channels (64).

The rest of this subsection will specify the design of TB in the proposed FCBFormer and how this varies from this definition of SSFormer. The improvements resulting from our changes are then demonstrated in the experimental section of this paper.

\subsubsection{Transformer encoder}
As in SSFormer, we used the PVTv2 \cite{pvtv2} for the image encoder in TB, pre-trained on ImageNet \cite{imagenet}. The variant of PVTv2 used is the B3 variant, which has 45.2M parameters. This model demonstrates exceptional feature extraction capabilities for dense prediction owing to its pyramid feature representation, contrasting with more traditional vision transformers which maintain the size of the spatial dimensions throughout the network, e.g. \cite{vit,deit,dpt}. Additionally, the model embeds the position of patches through zero padding and overlapping patch embedding via strided convolution, as opposed to adding explicit position embeddings to tokens, and for efficiency uses linear spatial reduction attention. On this element we do not deviate from the design of SSFormer.

\subsubsection{Improved progressive locality decoder (PLD+)}\label{newPLD}
We improve on the progressive locality decoder (PLD) introduced with SSFormer using the architecture shown in Fig. \ref{fig:FCBformer}b (PLD+), where we use residual blocks (RBs) (Fig. \ref{fig:FCBformer}f) to overcome identified limitations of the SSFormer's LE and SFA. These RBs take inspiration from the components of modern convolutional neural networks which have seen boosts in performance due to the incorporation of group normalisation \cite{groupnorm}, SiLU activation functions \cite{gelu}, and residual connections \cite{resnet}. We identified SSFormer's LE and SFA as being limited due to a lack of such modern elements, and a relatively low number of layers. As such, we modified these elements in FCBFormer to form the components of PLD+. The improvements resulting from these changes are shown through ablation tests in the experimental section of this paper.

As in SSFormer, the number of channels returned by the first convolutional layer in the LE blocks 64. Every subsequent layer, except channel-wise concatenation, then returns the same number of channels (64).

\subsection{Fully convolutional branch (FCB)}
We define the fully convolutional branch (FCB) (Fig. \ref{fig:FCBformer}c) as a composition of residual blocks (RBs), strided convolutional layers for downsampling, nearest neighbour interpolation for upsampling, and dense U-Net style skip connections. This design allows for the extraction of highly fused multi-scale features at full-size, which when fused with the important but coarse features extracted by the transformer branch (TB) allows for inference of full-size segmentation maps in the prediction head (PH).

Through the encoder of FCB, we increase the number of channels returned by each layer by a factor of 2 in the first convolutional layer of the first RB following the second and fourth downsampling layers. Through the decoder of FCB, we then decrease the number of channels returned by each layer by a factor of 2 in the first convolutional layer in the first RB after the second and fourth upsampling layers.

\subsection{Prediction head (PH)}
The prediction head (PH) (Fig. \ref{fig:FCBformer}d) takes a full-size tensor resulted from concatenating the up-sampled transformer branch (TB) output and the output from the fully convolutional branch (FCB). The PH predicts the segmentation map from important but coarse features extracted by TB by fusing them with the fine-grained features extracted by FCB. This approach for the combination of FCNs and transformers for dense prediction to the best of our knowledge has not been used before. As shown by our experiments, this approach is highly effective in polyp segmentation and indicates that FCNs and transformers operating in parallel prior to the fusion of features and pixel-wise prediction on the fused features is a powerful basis for dense prediction. Each layer of PH returns 64 channels, except the prediction layer which returns a single channel.

\section{Experiments}\label{sec:experiments}
To evaluate the performance of FCBFormer in polyp segmentation, we considered 2 popular open datasets, Kvasir-SEG \cite{kvasir}\footnote{Available: \url{https://datasets.simula.no/kvasir-seg/}} and CVC-ClinicDB \cite{cvc}\footnote{Available: \url{https://polyp.grand-challenge.org/CVCClinicDB/}}, and trained our models using the implementation detailed in Section \ref{implementation}. These datasets provide 1000/612 (Kvasir-SEG/CVC-ClinicDB) ground truth input-target pairs in total, with the samples in Kvasir-SEG varying in the size of the spatial dimensions while all samples in CVC-ClinicDB are of $288\times 384$ spatial dimensions. All images across both datasets contain polyps of varying morphology. These datasets have been used extensively in the developmentment of polyp segmentation models, and as such provide strong benchmarks for this assessment.

\subsection{Implementation details}\label{implementation}
We trained FCBFormer to predict binary segmentation maps of $h\times w$ spatial dimensions for RGB images resized to $h\times w$ spatial dimensions, where we set $h,w=352$ following the convention set by \cite{pranet,transfuse,ssformer}. We used PyTorch, and due to the aliasing issues with resizing images in such frameworks which have recently been brought to light \cite{aliasresizing}, we used anti-aliasing in our resizing of the images.  Both the images and segmentation maps were initially loaded in with a value range of $[0,1]$. We then used a random train/validation/test split of 80\%/10\%/10\% following the convention set by \cite{resunet++,doubleunet,pranet,msrfnet,ssformer}, and randomly augmented the training input-target pairs as they were loaded in during each epoch using: 1) a Gaussian blur with a $25\times 25$ kernel with a standard deviation uniformly sampled from $[0.001,2]$; 2) colour jitter with a brightness factor uniformly sampled from $[0.6,1.4]$, a contrast factor uniformly sampled from $[0.5,1.5]$, a saturation factor uniformly sampled from $[0.75,1.25]$, and a hue factor uniformly sampled from $[0.99,1.01]$; 3) horizontal and vertical flips each with a probability of 0.5; and 4) affine transforms with rotations of an angle sampled uniformly from $[-180\degree,180\degree]$, horizontal and vertical translations each of a magnitude sampled uniformly from $[-44,44]$, scaling of a magnitude sampled uniformly from $[0.5,1.5]$ and shearing of an angle sampled uniformly from $[-22.5\degree,22\degree]$. Out of these augmentations, 1) and 2) were applied only to the image, while the rest of the augmentations were applied consistently to both the image and the corresponding segmentation map. Following augmentation, the image RGB values were normalised to an interval of $[-1,1]$. We note that performance was achieved by resizing the segmentation maps used for training with bilinear interpolation without binarisation, however the values of the segmentation maps in the validation and test sets were binarised after resizing.

We then trained FCBFormer on the training set for each considered polyp segmentation dataset for 200 epochs using a batch size of 16 and the AdamW optimiser \cite{adamw} with an initial learning rate of 1e-4. The learning rate was then reduced by a factor of 2 when the performance (mDice) on the validation set did not improve over 10 epochs until reaching a minimum of 1e-6, and saved the model after each epoch if the performance (mDice) on the validation set improved. The loss function used was the sum of the binary cross entropy (BCE) loss and the Dice loss.

For comparison against alternative architectures, we also trained and evaluated a selection of well-established and state-of-the-art examples, which also predict full-size segmentation maps, on the same basis as FCBFormer, including: U-Net \cite{unet}, ResUNet \cite{resunet}, ResUNet++ \cite{resunet++}, PraNet \cite{pranet}, and MSRF-Net \cite{msrfnet}. This did not include SSFormer, as an official codebase has yet to be made available and the model by itself does not predict full-size segmentation maps. However, we considered our own implementation of SSFormer in an ablation study presented at the end of this section. To ensure these models were trained and evaluated in a consistent manner while ensuring training and inference was conducted as the authors intended, we used the official codebase\footnote{ResUNet++ code available: \url{https://github.com/DebeshJha/ResUNetPlusPlus}\\
PraNet code available: \url{https://github.com/DengPingFan/PraNet}\\
MSRF-Net code available:  \url{https://github.com/NoviceMAn-prog/MSRF-Net}} provided for each, where possible\footnote{For U-Net and ResUNet, we used the implementations built into the ResUnet++ codebase (available: \url{https://github.com/DebeshJha/ResUNetPlusPlus})} and modified this only to ensure that the models were trained and evaluated using data of $352\times 352$ spatial dimensions and that the same train/validation/test splits were used.

Some of the codebases for the existing models implement the respective model in TensorFlow/Keras, as opposed to PyTorch as is the case for FCBFormer. After observing slight variation in the results returned by the implementations of the considered metrics in these frameworks for the same inputs, we took steps to ensure a fair and balanced assessment. We therefore predicted the segmentation maps for each assessment within each respective codebase, after training, and saved the predictions. In a separate session using only Scikit-image, we then loaded in the targets for each assessment from source, resized to $352\times 352$ using bilinear interpolation, and binarised the result. The binary predictions were then loaded in, and we used the implementations of the metrics in Scikit-learn to obtain our results. Note that this was done for all models in each assessment.

\subsection{Evaluation}
\begin{figure}[htp]
\makebox[\textwidth][c]{\begin{tabular}{cccccccc}
  Input & Target & FF (ours) & PN \cite{pranet} & MN \cite{msrfnet} & R++ \cite{resunet++} & RU \cite{resunet} & UN \cite{unet} \\ 
 \includegraphics[width=1.5cm]{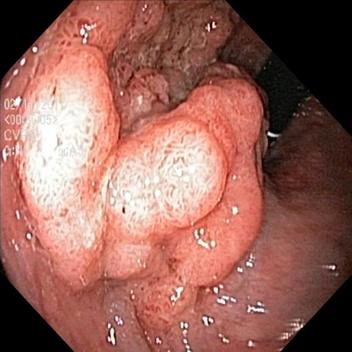}& \includegraphics[width=1.5cm]{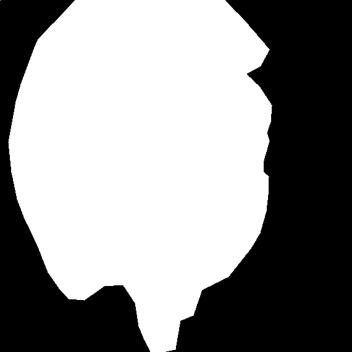} & \includegraphics[width=1.5cm]{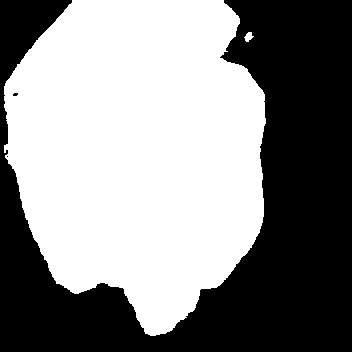} & \includegraphics[width=1.5cm]{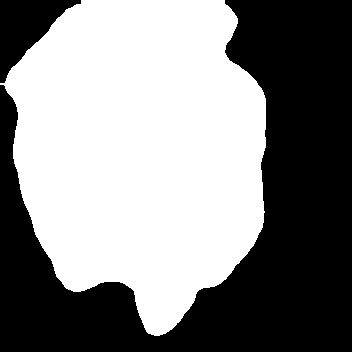} & \includegraphics[width=1.5cm]{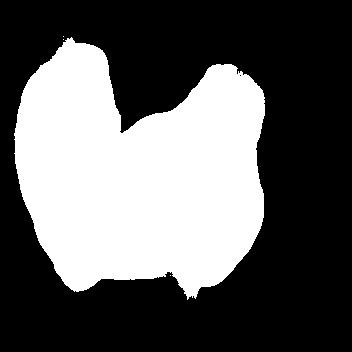} & \includegraphics[width=1.5cm]{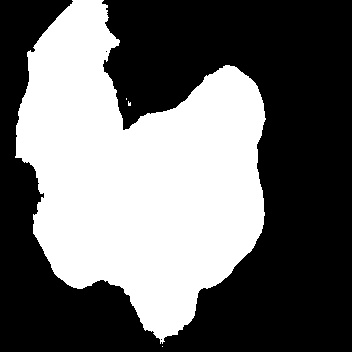} & \includegraphics[width=1.5cm]{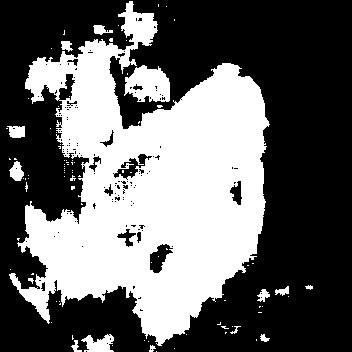} & \includegraphics[width=1.5cm]{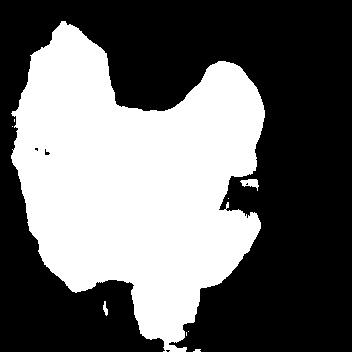} \\
  \includegraphics[width=1.5cm]{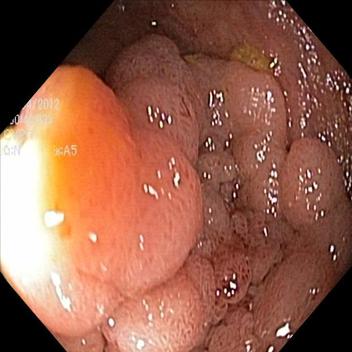}& \includegraphics[width=1.5cm]{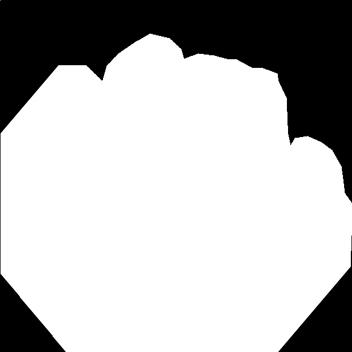} & \includegraphics[width=1.5cm]{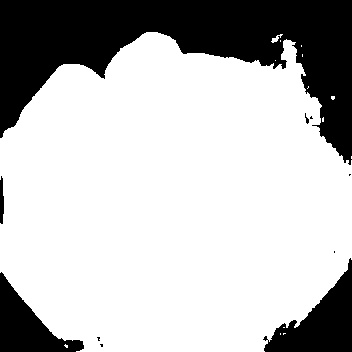} & \includegraphics[width=1.5cm]{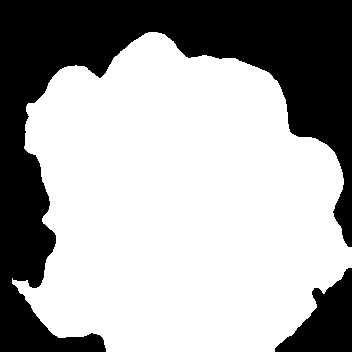} & \includegraphics[width=1.5cm]{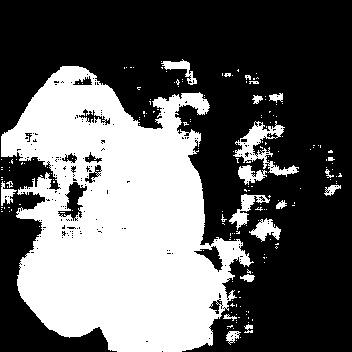} & \includegraphics[width=1.5cm]{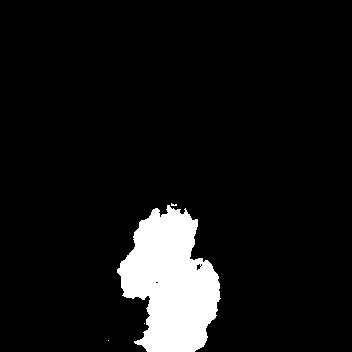} & \includegraphics[width=1.5cm]{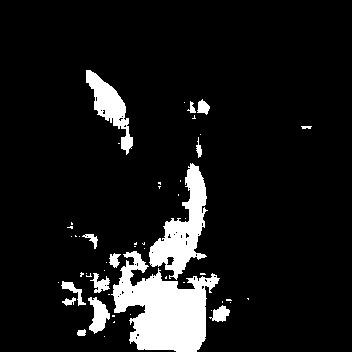} & \includegraphics[width=1.5cm]{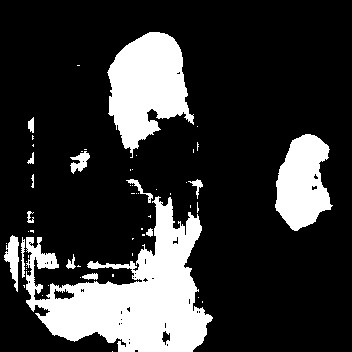} \\
\includegraphics[width=1.5cm]{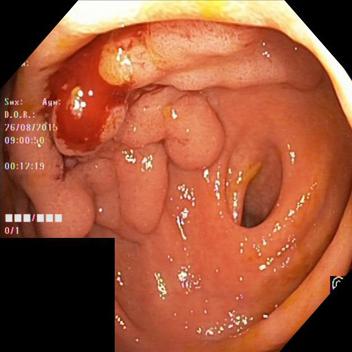}& \includegraphics[width=1.5cm]{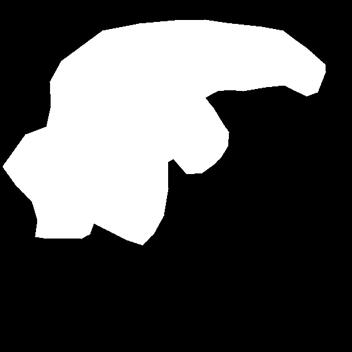} & \includegraphics[width=1.5cm]{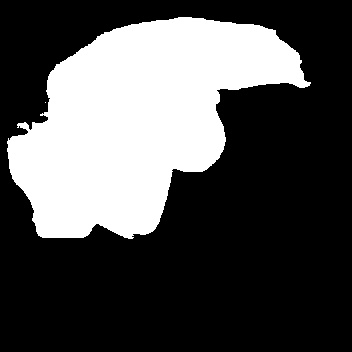} & \includegraphics[width=1.5cm]{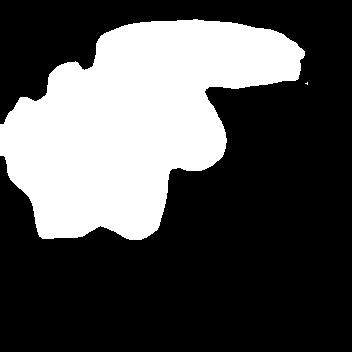} & \includegraphics[width=1.5cm]{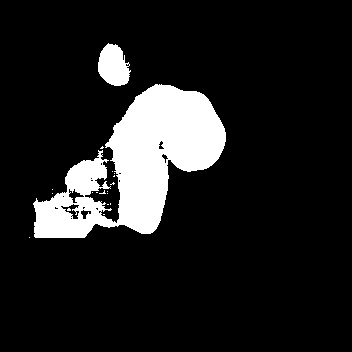} & \includegraphics[width=1.5cm]{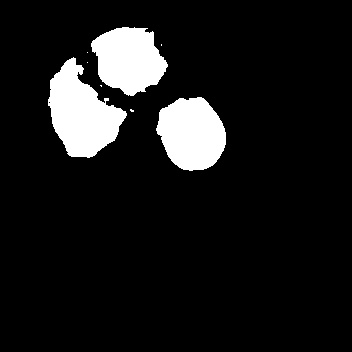} & \includegraphics[width=1.5cm]{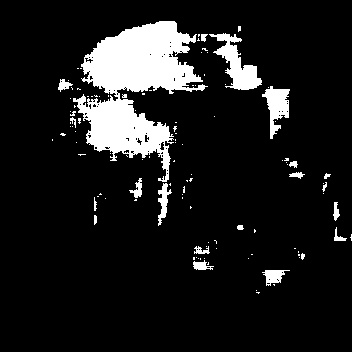} & \includegraphics[width=1.5cm]{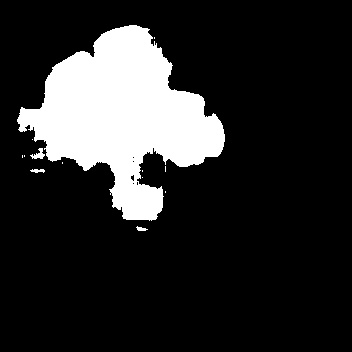} \\

\end{tabular}}
    \caption{Example inputs and targets from the Kvasir-SEG test set \cite{kvasir} and the predictions for FCBFormer and the considered existing architectures. FF is FCBFormer, PN is PraNet, MN is MSRF-Net, R++ is ResUNet++, RU is ResUNet, and UN is U-Net. Each model used for this was the variant trained on the Kvasir-SEG training set.}
    \label{fig:example_preds}
\end{figure}

\begin{figure}[htp]
\makebox[\textwidth][c]{\begin{tabular}{cccccc}
   Input & Target & TB output & FCB output & with FCB & without FCB \\
    \includegraphics[width=2cm]{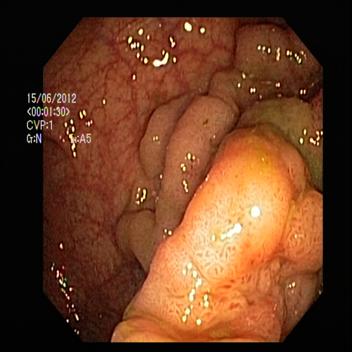} & \includegraphics[width=2cm]{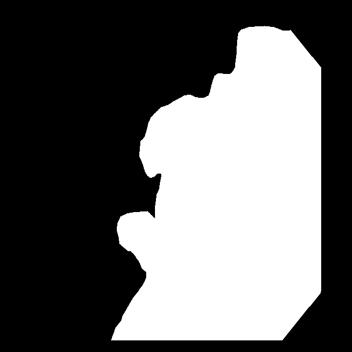}& \includegraphics[width=2cm]{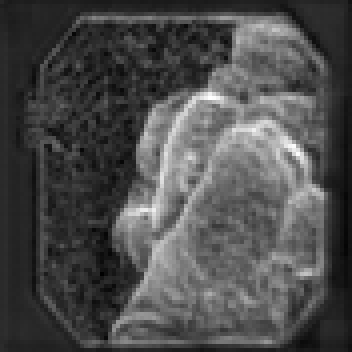} & \includegraphics[width=2cm]{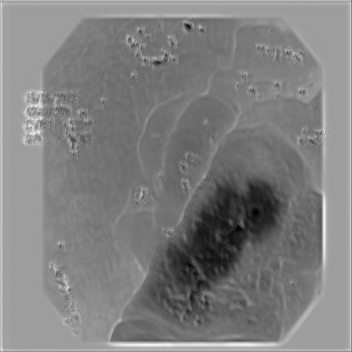} & \includegraphics[width=2cm]{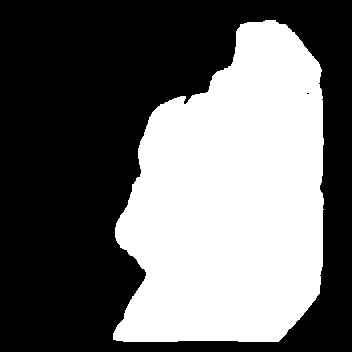} & \includegraphics[width=2cm]{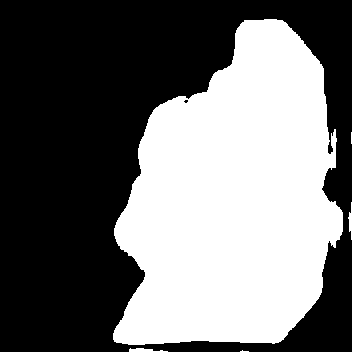}  \\
    \includegraphics[width=2cm]{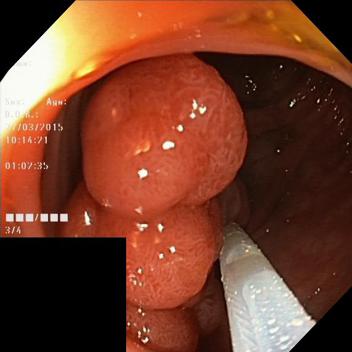} & \includegraphics[width=2cm]{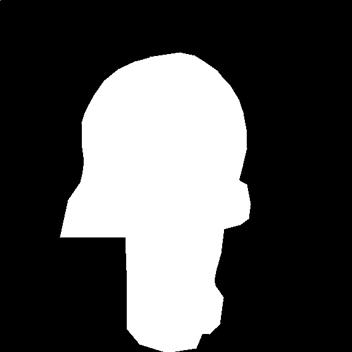}& \includegraphics[width=2cm]{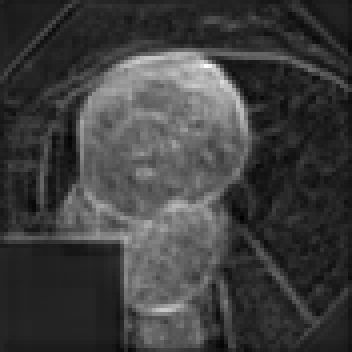} & \includegraphics[width=2cm]{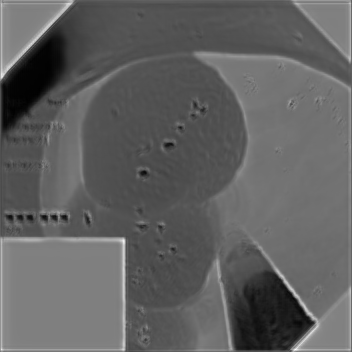} & \includegraphics[width=2cm]{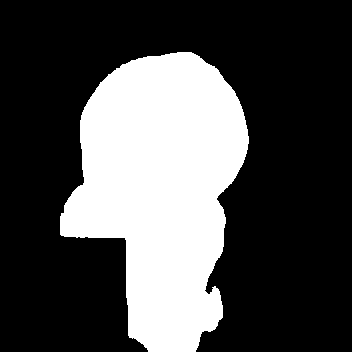} & \includegraphics[width=2cm]{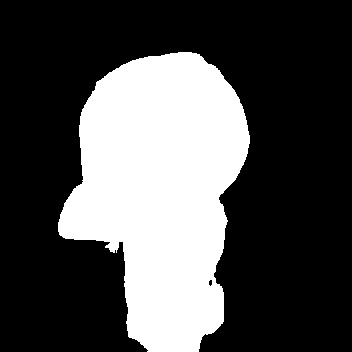}  \\
    \includegraphics[width=2cm]{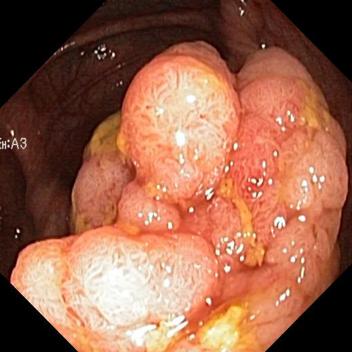} & \includegraphics[width=2cm]{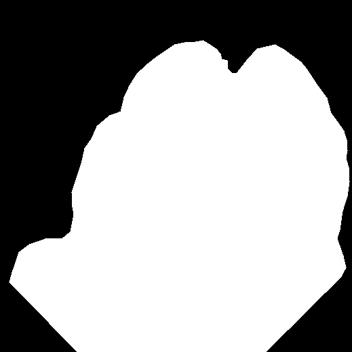}& \includegraphics[width=2cm]{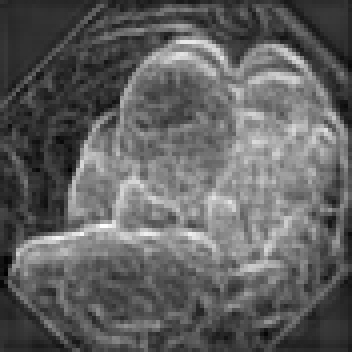} & \includegraphics[width=2cm]{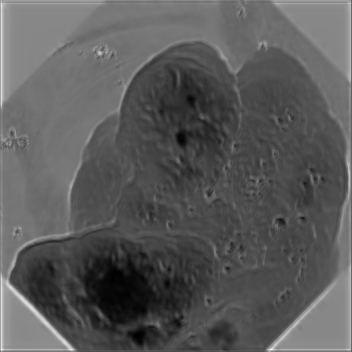} & \includegraphics[width=2cm]{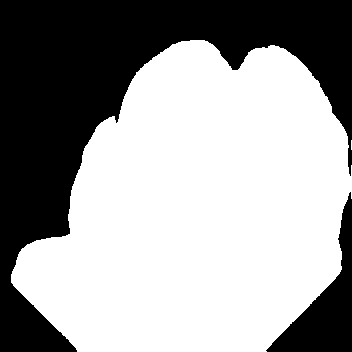} & \includegraphics[width=2cm]{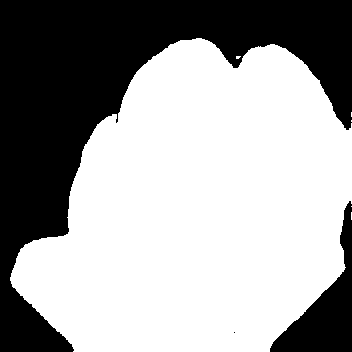}  \\
\end{tabular}}
    \caption{Visualisation of the features returned by TB and FCB (channel-wise average), and the with/without FCB predictions for examples from the Kvasir-SEG \cite{kvasir} test set.}
    \label{fig:example_feats}
\end{figure}

We present some example predictions for each model in Fig. \ref{fig:example_preds}. From this, it can be seen how FCBFormer predicts segmentation maps which are generally more consistent with the target than the segmentation maps computed by the existing models, and which demonstrate robustness to challenging morphology, highlighted by cases where the existing models are unable to represent the boundary well. This particular strength in segmenting polyps for which the boundary is less apparent is likely a result of the successful combination of the strengths of transformers and FCNs in FCBFormer, leading to the main structures of polyps being dealt with by the transformer branch (TB), while the fully convolutional branch (FCB) serves to ensure a reliable full-size boundary around this main structure. We demonstrate this in Fig. \ref{fig:example_feats}, where we show the features extracted by TB and FCB, and the predictions, for examples from the Kvasir-SEG \cite{kvasir} test set. The predictions are shown for the model with FCB, as defined, as well as for the model without FCB, where we concatenate the output of TB channel-wise with a tensor of 0's in place of the output of FCB. This reveals how the prediction head (PH) performs with and without the information provided by FCB, and in turn the role of FCB in assisting with the prediction. The most apparent function is that FCB highlights the edges of polyps, as well as the edges of features that may cause occlusions of polyps, such as other objects in the scene or the perimeter of the colonoscope view. This can then be seen to help provide a well-defined boundary, particularly when a polyp is near or partly occluded by such features.

\subsubsection{Primary evaluation}
For each dataset, we evaluated the performance of the models with respect to the mDice, mIoU, mPrecision, and mRecall metrics, where m indicates an average of the metric value over the test set. The results from these primary assessments are shown in Table \ref{primary}, which show that FCBFormer outperformed the existing models with respect to all metrics.

We note that for some of the previously proposed methods, we obtain worse results than has been reported in the  original papers, particularly MSRF-Net \cite{msrfnet}. This is potentially due to some of the implementations being optimised for spatial dimensions of size $256 \times 256$, as opposed to $352 \times 352$ as has been used here. This is supported by our retraining and evaluation of MSRF-Net \cite{msrfnet} with $256 \times 256$ input-targets, where we obtained similar results to those reported in the original paper. We therefore present the results originally reported by the authors of each model in Table \ref{authors}. Despite the potential differences in the experimental set up, it can be seen that FCBFormer consistently outperforms other models with respect to the observed mDice, one of the most important metrics out of those considered, and also outperforms other models with respect to mRecall on the Kvasir-SEG dataset \cite{kvasir}, and mPrecision on the CVC-ClinicDB dataset \cite{cvc}. FCBFormer can also be seen to perform competitively with respect to the mIoU.

\begin{table}[]
\caption{Results from our primary assessment.}\label{primary}
\makebox[\textwidth][c]{\begin{tabular}{|l|c|c|c|c|c|c|c|c|}
\hline
Dataset                                     & \multicolumn{4}{c|}{Kvasir-SEG \cite{kvasir}} & \multicolumn{4}{c|}{CVC-ClinicDB \cite{cvc}} \\ \hline
Metric                                      & mDice         & mIoU        & mPrec.        & mRec.        & mDice        & mIoU        & mPrec.        & mRec.        \\ \hline
U-Net \cite{unet} & 0.7821             & 0.8141           & 0.7241                & 0.8450             & 0.8464 & 0.7730 & 0.8496 & 0.8796             \\
ResUNet \cite{resunet} & 0.5133             & 0.3792           & 0.5937                & 0.5968             & 0.5221            & 0.4120           & 0.6151               & 0.5895             \\
ResUNet++ \cite{resunet++} & 0.8074             & 0.7231           & 0.8991                & 0.7874           & 0.5211            & 0.4126           & 0.5633               & 0.5693             \\
MSRF-Net \cite{msrfnet}    & 0.8586             & 0.7906           & 0.8933                & 0.8774             & 0.9198            & 0.8729           & 0.9222                & 0.9308             \\
PraNet \cite{pranet}              & 0.9011           & 0.8403                & 0.9034    & 0.9272         & 0.9358 & 0.8867  & 0.9370 & 0.93888             \\ \hline
FCBFormer (ours)                            & \textbf{0.9385}             & \textbf{0.8903}           & \textbf{0.9459}               & \textbf{0.9401}             & \textbf{0.9469}            & \textbf{0.9020}           & \textbf{0.9525}                & \textbf{0.9441}             \\ \hline
\end{tabular}}
\end{table}

\begin{table}[]
\caption{Results originally reported for existing models. Note that U-Net and ResUNet were not originally tested on polyp segmentation, and as such we present the results obtained by the authors of ResUNet++ \cite{resunet++} for these models. For ease of comparison, we include the results we obtained for FCBFormer in our primary assessment.}\label{authors}
\makebox[\textwidth][c]{\begin{tabular}{|l|c|c|c|c|c|c|c|c|}
\hline
Dataset                                     & \multicolumn{4}{c|}{Kvasir-SEG \cite{kvasir}} & \multicolumn{4}{c|}{CVC-ClinicDB \cite{cvc}} \\ \hline
Metric                                      & mDice         & mIoU        & mPrec.        & mRec.        & mDice        & mIoU        & mPrec.        & mRec.        \\ \hline
U-Net \cite{unet} & 0.7147 & 0.4334 & 0.9222 & 0.6306            & 0.6419 & 0.4711 & 0.6868 & 0.6756             \\
ResUNet \cite{resunet} & 0.5144 & 0.4364 & 0.7292 & 0.5041             & 0.4510 & 0.4570 & 0.5614 & 0.5775 \\
ResUNet++ \cite{resunet++} & 0.8133             & 0.7927           & 0.7064                & 0.8774             & 0.7955            & 0.7962                & 0.8785           & 0.7022             \\
MSRF-Net \cite{msrfnet}    & 0.9217             & \textbf{0.8914}           & \textbf{0.9666}                & 0.9198             & 0.9420            & \textbf{0.9043}           & 0.9427                & \textbf{0.9567}             \\
PraNet \cite{pranet}                 & 0.898 & 0.840                & -             & - & 0.899 & 0.849 & -  & -           \\ \hline
FCBFormer (ours)                            & \textbf{0.9385}             & 0.8903           & 0.9459               & \textbf{0.9401}             & \textbf{0.9469}            & 0.9020           & \textbf{0.9525}                & 0.9441             \\ \hline
\end{tabular}}
\end{table}

\subsubsection{Generalisability tests}
We also performed generalisability tests following the convention set by \cite{msrfnet,ssformer}. Using the same set of metrics, we evaluated the models trained on the Kvasir-SEG/CVC-ClinicDB training set on predictions for the full CVC-ClinicDB/Kvasir-SEG dataset. Such tests reveal how models perform with respect to a different distribution to that considered during training.

The results for the generalisability tests are given in Table \ref{generalisability}, where it can be seen that FCBFormer exhibits particular strength in dealing with images from a somewhat different distribution to those used for training, significantly outperforming the existing models with respect to most metrics. This is likely a result of the same strengths highlighted in the discussion of Fig. \ref{fig:example_preds}.

\begin{table}[]
\caption{Results from our generalisability tests.}\label{generalisability}
\makebox[\textwidth][c]{\begin{tabular}{|l|c|c|c|c|c|c|c|c|}
\hline
Training data                                     & \multicolumn{4}{c|}{Kvasir-SEG \cite{kvasir}} & \multicolumn{4}{c|}{CVC-ClinicDB \cite{cvc}} \\ \hline
Test data                                     & \multicolumn{4}{c|}{CVC-ClinicDB \cite{cvc}} & \multicolumn{4}{c|}{Kvasir-SEG \cite{kvasir}} \\ \hline
Metric                                      & mDice         & mIoU        & mPrec.        & mRec.        & mDice        & mIoU        & mPrec.        & mRec.        \\ \hline
U-Net \cite{unet} & 0.5940 & 0.5081 & 0.6937 & 0.6184             & 0.5292 & 0.4036 & 0.4613 & 0.8481             \\
ResUNet \cite{resunet} & 0.3359 & 0.2425 & 0.5048 & 0.3307            & 0.3344 & 0.2222 & 0.2618 & 0.8164             \\
ResUNet++ \cite{resunet++} & 0.5638 & 0.4750 & 0.7175 & 0.5908             & 0.3077 & 0.2048 & 0.3340 & 0.4778             \\
MSRF-Net \cite{msrfnet}    & 0.6238           & 0.5419                & 0.6621     & 0.7051 & 0.7296 & 0.6415 & 0.8162 & 0.7421             \\
PraNet \cite{pranet}    & 0.7912 & 0.7119 & 0.8152 & 0.8316 & 0.7950 & 0.7073 & 0.7687 & \textbf{0.9050}             \\ \hline
FCBFormer (ours)                            & \textbf{0.8735}             & \textbf{0.8038}           & \textbf{0.8995}                & \textbf{0.8876}             & \textbf{0.8848}            & \textbf{0.8214}           & \textbf{0.9354}                & 0.8754             \\ \hline
\end{tabular}}
\end{table}

As in our primary assessment, we also present results reported elsewhere. Similar generalisability tests were undertaken by the authors of MSRF-Net \cite{msrfnet}, leading to the results presented in Table \ref{authors2}. Again, we observe that FCBFormer outperforms other models with respect to most metrics. 

\begin{table}[]
\caption{Results from the generalisability tests conducted by the authors of MSRF-Net \cite{msrfnet}. Note, ResUNet \cite{resunet} was not included in these tests. For ease of comparison, we include the results we obtained for FCBFormer in our generalisability tests.}\label{authors2}
\makebox[\textwidth][c]{\begin{tabular}{|l|c|c|c|c|c|c|c|c|}
\hline
Training data                                     & \multicolumn{4}{c|}{Kvasir-SEG \cite{kvasir}} & \multicolumn{4}{c|}{CVC-ClinicDB \cite{cvc}} \\ \hline
Test data                                     & \multicolumn{4}{c|}{CVC-ClinicDB \cite{cvc}} & \multicolumn{4}{c|}{Kvasir-SEG \cite{kvasir}} \\ \hline
Metric                                      & mDice         & mIoU        & mPrec.        & mRec.        & mDice        & mIoU        & mPrec.        & mRec.        \\ \hline
U-Net \cite{unet} & 0.7172 & 0.6133 & 0.7986 & 0.7255             & 0.6222 & 0.4588 & 0.8133 & 0.5129             \\
ResUNet++ \cite{resunet++} & 0.5560 & 0.4542 & 0.6775 & 0.5795             & 0.5147 & 0.4082 & 0.7181 & 0.4860             \\
MSRF-Net \cite{msrfnet}    & 0.7921           & 0.6498                & 0.7000     & \textbf{0.9001} & 0.7575 & 0.6337 & 0.8314 & 0.7197             \\
PraNet \cite{pranet}    & 0.7225 & 0.6328 & 0.7888 & 0.7531 & 0.7293 & 0.6262 & 0.7623 & 0.8007             \\ \hline
FCBFormer (ours)                            & \textbf{0.8735}             & \textbf{0.8038}           & \textbf{0.8995}                & 0.8876             & \textbf{0.8848}            & \textbf{0.8214}           & \textbf{0.9354}                & \textbf{0.8754}             \\ \hline
\end{tabular}}
\end{table}

\subsubsection{Ablation study}
We also performed an ablation study, where we started from our implementation of SSFormer given in Fig. \ref{fig:SSformer}, since an official codebase has yet to be made available, and stepped towards FCBFormer. We refer to our implementation of SSFormer as SSFormer-I. This model was trained to predict segmentation maps of $\frac{h}{4}\times\frac{w}{4}$ spatial dimensions, and its performance in predicting full-size segmentation maps was then assessed by upsampling the predictions to $h\times w$ using bilinear interpolation then binarisation. We then removed the original prediction layer and used the resulting architecture as the transformer branch (TB) in FCBFormer, to reveal the benefits of our fully convolutional branch (FCB) and prediction head (PH) for full-size segmentation in isolation of the improved progressive locality decoder (PLD+), and we refer to this model as SSFormer-I+FCB. The additional performance of FCBFormer over SSFormer-I+FCB then reveals the benefits of PLD+. Note that SSFormer-I and SSFormer-I+FCB were both trained and evaluated on the same basis as FCBFormer and the other considered existing state-of-the-art architectures.

The results from this ablation study are given in Tables \ref{primary ablation} and \ref{generalisability ablation}, which indicate that: 1) there are significant benefits of FCB, as demonstrated by SSFormer-I+FCB outperforming SSFormer-I with respect to most metrics; and 2) there are generally benefits of PLD+, demonstrated by FCBFormer outperforming SSFormer-I+FCB on both experiments in the primary assessment and 1 out of 2 of the generalisability tests, with respect to most metrics.

\begin{table}[]
\caption{Results from the primary assessment in the ablation study. For ease of comparison, we include the results we obtained for FCBFormer in our primary assessment.}\label{primary ablation}
\makebox[\textwidth][c]{\begin{tabular}{|l|c|c|c|c|c|c|c|c|}
\hline
Dataset                                     & \multicolumn{4}{c|}{Kvasir-SEG \cite{kvasir}} & \multicolumn{4}{c|}{CVC-ClinicDB \cite{cvc}} \\ \hline
Metric                                      & mDice         & mIoU        & mPrec.        & mRec.        & mDice        & mIoU        & mPrec.        & mRec.        \\ \hline
SSFormer-I & 0.9196             & 0.8616           & 0.9316                & 0.9226             & 0.9318            & 0.8777           & 0.9409                & 0.9295             \\
SSFormer-I+FCB    & 0.9337             & 0.8850           & 0.9330                & \textbf{0.9482}             & 0.9410            & 0.8904           & \textbf{0.9556}                & 0.9307             \\ \hline
FCBFormer                            & \textbf{0.9385}             & \textbf{0.8903}           & \textbf{0.9459}               & 0.9401             & \textbf{0.9469}            & \textbf{0.9020}           & 0.9525                & \textbf{0.9441}             \\ \hline
\end{tabular}}
\end{table}

\begin{table}[]
\caption{Results from the generalisability test in the ablation study. For ease of comparison, we include the results we obtained for FCBFormer in our generalisability tests.}\label{generalisability ablation}
\makebox[\textwidth][c]{\begin{tabular}{|l|c|c|c|c|c|c|c|c|}
\hline
Training data                                     & \multicolumn{4}{c|}{Kvasir-SEG \cite{kvasir}} & \multicolumn{4}{c|}{CVC-ClinicDB \cite{cvc}} \\ \hline
Test data                                 & \multicolumn{4}{c|}{CVC-ClinicDB \cite{cvc}} & \multicolumn{4}{c|}{Kvasir-SEG \cite{kvasir}} \\ \hline
Metric                                      & mDice         & mIoU        & mPrec.        & mRec.        & mDice        & mIoU        & mPrec.        & mRec.        \\ \hline
SSFormer-I & 0.8611             & 0.7813           & 0.8904                & 0.8702             & 0.8691            & 0.7986           & 0.9178                & 0.8631             \\
SSFormer-I+FCB    & \textbf{0.8754}             & 0.8059           & \textbf{0.8935}                & \textbf{0.8963}             & 0.8704            & 0.7993           & 0.9280                & 0.8557         \\\hline
FCBFormer                            & 0.8735             & \textbf{0.8038}           & 0.8995                & 0.8876             & \textbf{0.8848}            & \textbf{0.8214}           & \textbf{0.9354}                & \textbf{0.8755}             \\ \hline
\end{tabular}}
\end{table}

\section{Conclusion}\label{sec:conclusion}
In this paper, we introduced the FCBFormer, a novel architecture for the segmentation of polyps in colonoscopy images which successfully combines the strengths of transformers and fully convolutional networks (FCNs) in dense prediction. Through our experiments, we demonstrated the models state-of-the-art performance in this task and how it outperforms existing models with respect to several popular metrics, and highlighted its particular strengths in generalisability and in dealing with polyps of challenging morphology. This work therefore represents another advancement in the automated processing of colonoscopy images, which should aid in the necessary improvement of lesion detection rates and classification.

Additionally, this work has interesting implications for the understanding of neural network architectures for dense prediction. The method combines the strengths of transformers and FCNs, by running a model of each type in parallel and concatenating the outputs for processing by a prediction head (PH). To the best of our knowledge, this method has not been used before, and its strengths indicate that there is still a great deal to understand about these different architecture types and the basis on which they can be combined for optimal performance. Further work should therefore explore this in more depth, by evaluating variants of the model and performing further ablation studies. We will also consider further investigation of dataset augmentation for this task, where we expect the random augmentation of segmentation masks to aid in overcoming variability in the targets produced by different annotators.

\subsubsection{Acknowledgements}
This work was supported by the Science and Technology Facilities Council grant number ST/S005404/1.

Discretionary time allocation on DiRAC Tursa HPC was also used for methods development.

\bibliographystyle{splncs04}
\bibliography{Paper.bib}

\end{document}